\documentclass{emulateapj}
\usepackage{graphicx, amsmath, amssymb, amsthm}
\usepackage{natbib}
\usepackage{apjfonts}

\begin{document}

\title{Mildly relativistic X-ray transient 080109 and SN\,2008D:\\ Towards a continuum from energetic GRB/XRF to ordinary Ibc SN}

\author{D. Xu\altaffilmark{1}, Y. C. Zou\altaffilmark{2,3}, and Y. Z. Fan \altaffilmark{4,5}}
\altaffiltext{1}{Dark Cosmology Centre, Niels Bohr Institute, University of Copenhagen,
Juliane Maries Vej 30, 2100, Copenhagen, Denmark} \altaffiltext{2}{The Racah Inst. of Physics,
Hebrew University, Jerusalem 91904, Israel} \altaffiltext{3}{Department of Physics, Huazhong
University of Science and Technology, 430074 Wuhan, China} \altaffiltext{4}{Neils Bohr
International Academy, Niels Bohr Institute, University of Copenhagen, Blegdamsvej 17, DK-2100
Copenhagen, Denmark}\altaffiltext{5}{Purple Mountain Observatory, Chinese Academy of Sciences,
Nanjing 210008, China} \email{yizhong@nbi.dk (YZF)}

\begin{abstract}
We analyze the hitherto available space-based X-ray data as well as ground-based optical data
of the X-ray transient 080109/SN\,2008D. From the data we suggest that ({\it i}) The initial
transient ($\lesssim 800$ sec) is attributed to the reverse shock emission of a mildly
relativistic ($\Gamma \sim$ a few) outflow stalled by the dense stellar wind. ({\it ii}) The
subsequent X-ray afterglow ($\lesssim 2\times 10^4$ sec) can be ascribed to the forward shock
emission of the outflow, with a kinetic energy $\sim 10^{46}$ erg, when sweeping up the
stellar wind medium. ({\it iii}) The late X-ray flattening ($\gtrsim 2\times 10^4$ sec) is
powered by the fastest non-decelerated
component of SN\,2008D's ejecta. 
({\it iv}) The local event rate of X-ray transient has a lower limit of $\sim 1.6\times
10^4~{\rm yr^{-1}~Gpc^{-3}}$, indicating a vast majority of X-ray transients have a wide
opening angle of $\gtrsim 100^\circ$. ({\it v}) Transient 080109/SN\,2008D indicates a
continuum from GRB-SN to under-luminous GRB-/XRF-SN to X-ray transient-SN and to ordinary Ibc
SN (if not every Ibc SN has a relativistic jet), as shown in Figure 2 of this {\it Letter}.
\end{abstract}


\section{Introduction}\label{sec:Into}
During the past decade, long-duration ($\gtrsim 2$sec) $\gamma-$ray bursts (GRBs), including
the subclass of X-ray flashes (XRFs), have been found (1) to be driven by the core-collapse of
massive stars \cite{Woosley93}; thus (2) to be associated with a rare variety ($\sim1\%$) of
type Ibc supernovae (SNe), the so-called hypernovae (HN)
\cite{Galama98,Hjorth03,Stanek03,Male04,Sollerman06,Camp06} (but also see Fynbo et al. 2006);
and (3) in general to be hosted by the star-forming dwarf galaxies with low metallicity
\cite{Fyn03,Fru06,Stanek06}. Though the association of GRB/XRF and Ibc SN has been pinned
down, what channels make a dying star to produce a GRB or an XRF, and not just a Ibc SN, is
still unclear. The progenitor's mass, metallicity, angular momentum, and the configuration and
strength of its internal magnetic field play important roles for the generation of GRBs/XRFs
and ordinary Ibc SNe.

The serendipitous discovery of the X-ray transient 080109/SN\,2008D may shed light on filling
in this gap between energetic GRBs/XRFs and ordinary Ibc SNe. We will analyze space- and
ground-based data of this transient and SN, focusing on X-ray/radio data because
observationally they trace the fastest component of the transient/SN outflow while optical
data trace the slower SN ejecta (e.g., Soderberg et al. 2006).

\section{Swift Observations and Data Analysis}
During {\it Swift}/XRT follow-up observations of Ib SN 2007uy beginning at 13:32:49 UT on Jan
9, 2008, an X-ray transient (Transient hereafter) was identified and reported on Jan 10.58
\cite{BerSod08a}. X-ray emission was already underway at time of trigger. Both Transient
080109 and SN 2007uy are in the same host galaxy, NGC2770, at $z=0.0065$. The object was
within the {\it Swift}/BAT field of view for approximately 30 minutes prior to the XRT
observation but never triggered BAT \cite{Burrows08}.

Since trigger, the Transient was observed to rise to a maximum flux about 65 seconds, and then
subsequently decay until the end of the first orbit, roughly 500 seconds afterwards. We
reduced the XRT data in a standard way using the Swift analysis software (HEAsoft 6.4) and
calibration data. The contamination by a source close to the Transient and pile-up have been
corrected.

A general spectral softening was seen during the first orbit. Taking a mean spectrum, the data
can be well fitted by either an absorbed power-law $\Gamma=2.3\pm0.2$ and a column density
$N_H=7.6^{+1.4}_{-1.2}\times10^{21}\,{\rm cm^{-2}}$ ($\chi _{dof}^2 =15.1/20$) with respect to
the Galactic number $1.7\times10^{20}\,{\rm cm^2}$ or by an absorbed blackbody spectrum with
$kT = 0.73\pm0.05~ {\rm keV}$ in the restframe ($\chi _{dof}^2 =24.2/20$). For the late X-ray
data, the spectral index cannot be well fitted due to the limited photon numbers but $\Gamma
\sim 2.1$ is acceptable.

UVOT marginally detected an optical counterpart to the X-ray transient on Jan 9 in the B
(3.4$\sigma$) and U (3.0$\sigma$) filters. After a data gap of 2 days, the source brightened
and showed up in both optical and UV. The source then faded until day $\sim3.5$ and
subsequently brightened again, confirming the onset of SN\,2008D (Page et al. 2008).

\begin{figure}
\centerline{\includegraphics[width=9cm, height=10cm, angle=0]{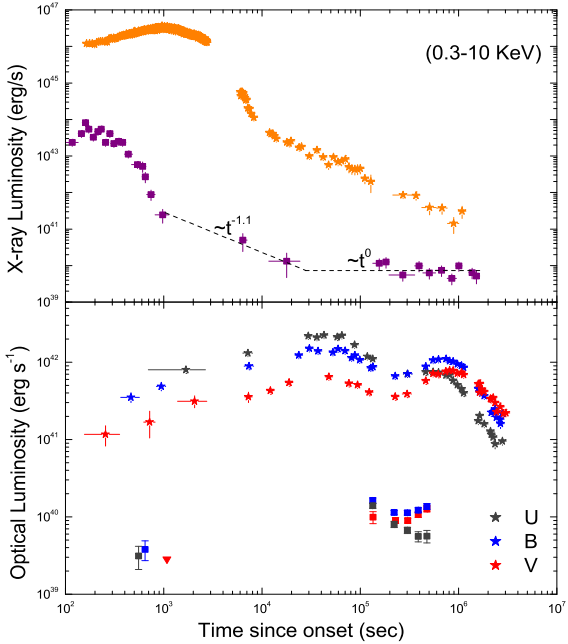}} \caption{Comparison of
X-ray transient\,080109/SN\,2008D (squares) and XRF 060218/SN 2006aj (stars). Data of
XRF\,060218/SN\,2006aj are taken from Campana et al. (2006). {\it Upper}: Temporal evolution
of the X-ray luminosity in 0.3-10 keV. {\it Lower}: The U (grey), B (blue), and V (red)
lightcurves for two events. Data of Transient 080109 have not been corrected for extinction.
\label{afterglow}}
\end{figure}

\section{Early spectroscopy and its evolution}
Two spectra were obtained at the ESO VLT equipped with FORS2 starting 07:17 UT on Jan 11. As
pointed in Malesani et al. (2008), the overall spectral shape rules out a significant
non-thermal afterglow component, but a SN one instead. The presence of broad features reveal
SN\,2008D's emergence, but the features are not as broad as in the earliest spectra of
GRB/XRF-associated SNe such as SN\,1998bw and SN\,2006aj.

SN\,2008D is distinguished for its apparent Ic$\rightarrow$Ib spectroscopic evolution. It was
classified as peculiar type-Ic on Jan 11, and then as a type-Ic with possibly some He as seen
in NIR spectra on Jan 13-15, and later to a type-Ib on Jan 21 (Modjaz et al. 2008 and
reference therein). This evolution is reminiscent of SN\,2005bf (Folatelli et al. 2006).

\section{Interpretation of the follow-ups}
\subsection{Real onset time and Spectrum for the first orbit}
Though it's impossible to know the exact onset time of the Transient, we may get a rough
estimate about this time-back shift. First, non-trigger of BAT approximately 30 minutes prior
to the XRT trigger and association with SN\,2008D give us confidence that Transient 080109 is
a dwarf outburst compared with previous bursts featuring low $\nu F_\nu$ peak energy such as
XRF\,060218. Second, the existence of a main outburst 30 minutes earlier would make the first
orbit lightcurve look like a very sharp flare with a decay index $\sim 20$ covering thee order
of magnitudes in luminosity. Its profile is not similar to the Fast-Rise-Exponential-Decay one
typical for GRB/XRF. Considering the above two factors, a back shift of tens to $\sim 200$ sec
is generally acceptable and doesn't affect the temporal decay laws after 1000 sec. In this
work, we adopt a back shift of $\sim$100 sec.

Should a shock break-out be responsible for the first orbit observation, according to the
equation $L = \Omega (\gamma ^2 ct)^2 \sigma T_{rest}^4$, where $L\sim10^{43} {\rm
erg\,s^{-1}}$ is the isotropic luminosity, $\Omega$ is the solid angle for the outburst,
$\gamma$ is the Lorentz factor for the outburst with respect to the observer, $\sigma$ is the
Stefan-Boltzmann constant, $T_{rest}$ is the temperature in the restframe, we then have
\[
T_{obs} \sim \frac{{1.6}}{{\Omega ^{1/4} t^{1/2} \gamma ^2 }} {\rm keV},
\]
where $T_{obs}$ is the measured temperature. To match the measured $T=0.73$ keV, a jet-like
outflow with $\Omega<10^{-6}$ is needed, which renders the shock break-out model unacceptable
in this event.

\subsection{The X-ray transient powered by the reverse shock of the mildly relativistic outflow }
The Transient lightcurve is smooth and decays as $t^{-3.7}$. The smoothness largely disfavors
that these X-ray photons are powered by internal shocks, and the steep decline rules out that
the transient is from the forward shock of an outflow.

We consider a mild-relativistic ($\Gamma_{\rm i} \sim$ 3) outflow
with a luminosity $L_{\rm m} \sim 10^{44}~\Omega_o^{-1}~{\rm
erg~s^{-1}}$ decelerated by the stellar wind medium, where
$\Omega_o$ is the solid angle of the initial transient outflow. The
density profile of the stellar wind is $n=3\times 10^{35}A_*
R^{-2}$, where the free wind parameter
$A_*=[\dot{M}/10^{-5}M_\odot~{\rm yr^{-1}}][v_w/(10^8~{\rm cm}~{\rm
s^{-1}})]$, $\dot{M}$ is the mass loss rate of the progenitor, and
$v_w$ is the velocity of the stellar wind \cite{LC99}. Because of
the low luminosity of the outflow while the high density of the
stellar wind, the forward-shocked material would move
sub-relativistically (i.e., $\Gamma_{\rm fr}\sim 1$) while the
reverse shock is mild-relativistic. The reverse shock region is thus
very hot and has a very large sideways expansion velocity of $\sim
c$, the speed of light. As a result, after the reverse shock crosses
the outflow, the shocked outflow will have a large solid angle
$\Omega\gg \Omega_o$ if $\Omega_o$ is small. So for simplicity we
treat the outflow as isotropic.

Now we estimate the synchrotron radiation of the reverse shock at a
distance $R_{\rm r}\sim cT_{90} \sim 10^{13}$ cm. As usual, we
assume $\epsilon_{\rm e}$ and $\epsilon_{\rm B}$ fractions of the
shock energy given to the electrons and magnetic field, respectively
\cite{spn98}. The minimum Lorentz factor of the reverse shock
accelerated electrons is $\gamma_{\rm m,r} \sim (\Gamma_{\rm
i}-1)\epsilon_{\rm e} (p-2)m_{\rm p}/[(p-1)m_{\rm e}]\sim 100$,
where $p\sim 2.2$ is the power-law index of the electrons, and the
magnetic filed generated in the reverse shock region is\footnote{The
convention $Q_x=Q/10^x$ has been adopted in cgs.} $B_{\rm r} \sim
[2(\epsilon_{\rm B}/\epsilon_{\rm e})L_{\rm m}/(R_{\rm r}^2
c)]^{1/2} \sim 3\times 10^5~{\rm Gauss}~ (\epsilon_{\rm
B}/\epsilon_{\rm e})^{1/2} L_{\rm m,44}^{1/2} R_{\rm r,13}^{-1}$.
The typical synchrotron radiation frequency $\nu_{\rm m,r} \sim
2.8\times 10^6~{\rm Hz}~\gamma_{\rm m,r}^2 B_{\rm r} \sim 3\times
10^{16}$ Hz, which is in the soft X-ray band and matches the
observation. The cooling Lorentz factor is $\gamma_{\rm c,r} \sim
7.7\times 10^8/(B_{\rm r}^2 t)\ll \gamma_{\rm m,r}$, so the reverse
shock is in the fast cooling phase. At a first glimpse, one might
interpret the $\sim t^{-3.7}$ decline as the high latitude emission
of the reverse shock. But this requires either a sub-relativistic
outflow with a very sharp energy distribution from the outflow
center to the outflow side or a $\Gamma_{\rm fr}\geq$ a few. The
validity of the former option is hard to estimate. The latter is
also difficult to function because $\Gamma_{\rm fr} \approx 1.1
L_{\rm m,44}^{1/4}A_{*,-1}^{-1/4}$. A $\Gamma_{\rm fr}\sim 3$
requires $L_{\rm m} \sim 10^{47}~{\rm erg~s^{-1}}~ A_*$, which is
too high to match the observation.

In this work we interpret the steep X-ray decline as the dimmer and dimmer reverse shock
emission powered by the weaker and weaker outflow. Within this scenario, the outflow is very
likely to be mildly relativistic. If $\Gamma_{\rm i} \sim $ tens$-$hundreds, $\gamma_{\rm
m,r}\sim 10^3-10^4$ and $\nu_{\rm m,r}$ would be $\sim 5-500$ keV. As a result, the XRT
spectrum should be $\propto \nu^{-0.6}$, which is inconsistent with the observed value. A
marginally relativistic ($\Gamma_{\rm i} \sim 1.2$) outflow model is also disfavored because
the reverse shock would be very weak. With typical shock parameters, the emission of such a
weak reverse shock can not peak at soft X-ray band and is likely to be outshone by the forward
shock X-ray emission.

\subsection{The early X-ray afterglow ($\lesssim2\times 10^4$ sec) powered by the forward shock of the Transient outflow}
We calculate the synchrotron radiation of the sub-relativistic
forward shock when it sweeps up the surrounding stellar wind medium.
The minimum Lorentz factor of the shocked electrons, the magnetic
field and the cooling Lorentz factor are $\gamma_{\rm m}\approx
16~C_{\rm p}\beta^2 \epsilon_{\rm e,-1}$, $B\approx 15~{\rm
Gauss}~\beta R_{15}^{-1}\epsilon_{\rm B,-1}^{1/2}A_{*,-1}^{1/2}$,
and $\gamma_{\rm c}\approx 34~\beta^{-2}R_{15}^2 \epsilon_{\rm
B,-1}^{-1}A_{*,-1}^{-1}t_5^{-1}$, respectively, where $C_{\rm
p}\equiv 6(p-2)/(p-1)$. The maximum specific flux, the typical
synchrotron radiation frequency and the cooling frequency
\cite{spn98} are respectively given by
\begin{equation}
F_{\nu, \rm max}
\approx 1~{\rm Jy}~\beta \epsilon_{\rm B,-1}^{1/2}A_{*,-1}^{3/2}
D_{\rm L,25.9}^{-2}, \label{eq:F_numax}
\end{equation}
\begin{equation}
\nu_{\rm m} \approx 1.1\times 10^{10}~{\rm Hz}~\beta^5 C_{\rm p}^2
R_{15}^{-1} \epsilon_{\rm e,-1}^2 \epsilon_{\rm
B,-1}^{1/2}A_{*,-1}^{1/2}, \label{eq:nu_m}
\end{equation}
\begin{equation}
\nu_{\rm c} \approx 4.8\times 10^{10}~{\rm Hz}~\beta^{-3} R_{15}^{3}
\epsilon_{\rm B,-1}^{-3/2}  A_{*,-1}^{-3/2}t_5^{-2}, \label{eq:nu_c}
\end{equation}
where $D_{\rm L}$ is the luminosity distance of the source. So the X-ray flux can be estimated
as
\begin{equation}
\begin{array}{l}
F_{\nu_{\rm X}}=F_{\nu,\rm max} \nu_{\rm c}^{1/2}\nu_{\rm
m}^{(p-1)/2}\nu_{\rm
X}^{-p/2}\\
=4.6\times 10^{-2}~{\rm \mu Jy}~ \nu_{\rm
X,17}^{-p/2}~\beta^{(5p-6)/2}\epsilon_{\rm e,-1}^{p-1} \epsilon_{\rm
B,-1}^{(p-2)/4}\\A_{*,-1}^{(p+2)/4}D_{\rm L,25.9}^{-2}C_{\rm
p}^{p-1}R_{15}^{(4-p)/2}t_5^{-1}. \label{eq:F_x}
\end{array}
\end{equation}
If $\beta \sim $const., i.e., the outflow hasn't been decelerated
significantly, we have $R \approx \beta c t$ and $F_{\nu_{\rm X}}
\propto t^{(2-p)/2}$. The decline is thus too shallow to be
consistent with the detected $\propto t^{-1.1}$ for the early X-ray
afterglow. We then consider an alternative in which the outflow with
a energy distribution $E(\geq \beta \Gamma) \propto (\beta
\Gamma)^{-k}$ has entered the Sedov regime, thus we have $\beta
\propto t^{-{1\over 3+k}}$ and $R={3+k \over 2+k}\beta c t\propto
t^{2+k \over 3+k}$. Accordingly, Eqs.
(\ref{eq:F_numax}-\ref{eq:nu_c}) read $F_{\nu, \rm max}\propto
t^{-{1\over 3+k}}$,$\nu_{\rm m} \propto t^{-{7+k\over 3+k}}$, and
$\nu_{\rm c}\propto t$, respectively. The light curves are of
\begin{eqnarray}
F_\nu \propto ~\left\{%
\begin{array}{ll}
t^{4+k\over 3(3+k)} & {\rm for~~ \nu<\nu_m<\nu_c}, \\
  t^{-{1\over 3+k}[1+{(p-1)(7+k) \over 2}]} & {\rm for~~ \nu_m<\nu<\nu_c}, \\
    t^{-{1\over 3+k}[1+{(p-1)(7+k) \over 2}]+{1\over 2}} & {\rm for~~ \nu>\max\{\nu_c,\nu_m\}}.\\
\end{array}%
\right. \label{eq:LC}
\end{eqnarray}
So the early X-ray afterglow decline $F_{\nu_{\rm X}}\propto
t^{-1.1}$ suggests a very small $k\sim 0.4$ for $p\sim 2.2$,
implying that the outflow almost has a very flat energy distribution
(i.e., the Transient ejecta is likely expanding with a single bulk
Lorentz factor). So we assume $E_{\rm _{tran}} \approx 4\pi \beta^2
R^3 n m_{\rm p}c^2$, which yields
\begin{equation}
\beta \sim 0.23 ~E_{_{\rm tran},46.5}^{1/3} A_*^{-1/3}t_4^{-1/3}.
\label{eq:beta_limit}
\end{equation}
Eq.(\ref{eq:F_x}) thus reduces to
\begin{equation}
F_{\nu_{\rm X}} \sim 0.04~{\rm \mu Jy}~\epsilon_{\rm e,-1}^{1.2}
E_{_{\rm tran},46.5}^{1.1} t_4^{-1.2},
\end{equation}
which is consistent with the XRT flux $\sim 0.03~{\rm \mu Jy}$ at
$10^{17}$ Hz at $t \sim 10^{4}$ sec (see Fig.1). The outflow energy
inferred above is $E_{_{\rm tran}} \sim 3\times 10^{46}$ erg, which
is larger than the isotropic energy of the X-ray transient by a
factor of 10 and is reasonable.

Till here we have shown that a mild-relativistic outflow with an
energy $\sim 3\times 10^{46}$ erg can account for the Transient and
the early X-ray afterglow self-consistently, which implies that
there was no energetic outburst before the X-ray transient. Should
it happen, there would be a bright X-ray afterglow component, which
actually could outshine the current data. On the other hand, an
earlier outburst would sweep up the stellar wind medium and leave a
very low density bubble. The Transient outflow thus cannot get
decelerated effectively and cannot account for the following X-ray
afterglow data.

\subsection{The late X-ray afterglow ($\gtrsim2\times10^4$ sec) powered by the supernova shock}
After $\gtrsim2\times10^4$ sec, the X-ray lightcurve gets
flattening. We interpret this flattening as the shock emission of
the fastest component of the SN ejecta, which moves with a velocity
$\sim 0.2$ c (see Eq. (\ref{eq:beta_limit}) for the limit).

As shown in Eq. (\ref{eq:F_x}), if the SN fastest component is
energetic enough that hasn't got decelerated significantly in a
timescale $\sim 10^6$ sec or even longer, we have
\[F_{\nu_{\rm X}} \propto t^{(2-p)/2}\sim t^{-0.1},\] which is consistent with the observed
flattening (see Fig.1).

The observed X-ray flux at $\sim 10^{17}$ Hz at $t\sim 10^6$ s is
$\sim 0.02~{\rm \mu Jy}$, which requires $\beta \approx
0.1^{1/(2p-1)}\epsilon_{\rm
e,-1}^{(1-p)/(2p-1)}A_{*,-1}^{-(p+2)/[4(2p-1)]}$. A reasonable
choice of $A_{*} \sim 1$, $\epsilon_{\rm e,-1} \sim 1$, and $\beta
\sim 0.2$ leads to that the total energy of the SN fastest component
is no less than
\begin{eqnarray}
E_{\rm _{SN}}(\beta\geq 0.2) &\sim & 4\pi n R^3 \beta^2 m_p c^2
\nonumber\\
&\sim & 3\times 10^{48}~{\rm erg}~A_*^{{5(p-2) \over
4(2p-1)}}\epsilon_{\rm e,-1}^{3(1-p)\over 2p-1} t_{6}.
\end{eqnarray}

For comparison, we plot in Figure 2 the identified energy
distribution of the outflows associated with XRTr 080109/SN 2008D,
together with those of ordinary Ic SNe and the hypernovae associated
with energetic GRBs/XRFs. At first glimpse, the main difference
between the ordinary Ic SNe and the GRB/XRF-associated ones is the
energy of the (mild-)relativistic outflow.

\begin{figure}
\centerline{\includegraphics[width=11cm, height=12cm, angle=0]{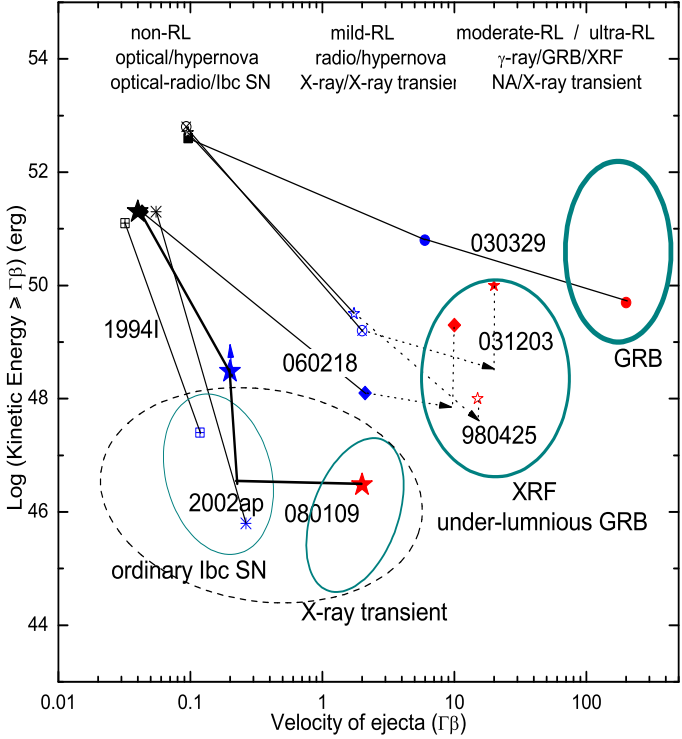}} \caption{Energy
distribution for Transient 080109-SN\,2008D as well as for GRB-HN, under-luminous GRB-/XRF-HN,
and ordinary Ibc SNe. RL and NA represent ``relativistic'' and ``not available'',
respectively. Part of data from Soderberg et al. (2006) and Kaneko et al. (2006). Sudden drop
of the energy distribution in GRB 031203 and XRF 060218 after the prompt emission might be due
to the geometry correction and/or a high GRB/XRF efficiency. Transient 080109/SN\,2008D marks
a transition between populations of ordinary Ibc SNe and under-luminous GRB-/XRF-HN. We
caution that X-ray transients may account for a majority of Ibc SN events. \label{continuum}}
\end{figure}

\subsection{The radio afterglow: the supernova shock model}
Radio emission below the self-absorption frequency, $\nu_{\rm a}$,
would be suppressed significantly. Through the standard treatment
\cite{RL79}, for $\nu_{\rm a}<\nu_{\rm m}<\nu_{\rm c}$, we have $
\nu_{\rm a} \approx 3.9\times 10^{12}~{\rm Hz}~
\beta^{-13/5}t_5^{-1}\epsilon_{\rm e,-1}^{-1}\epsilon_{\rm
B,-1}^{1/5}A_{*,-1}^{4/5}$. While for $\nu_{\rm m}<\nu_{\rm
a}<\nu_{\rm c}$, we have
\begin{equation}
\nu_{\rm a} \approx 2.6\times 10^{11}~{\rm Hz}~  \beta^{4p-6 \over
p+4}t_5^{-1}\epsilon_{\rm e,-1}^{2-p \over p+4}\epsilon_{\rm
B,-1}^{p+2 \over 2(p+4)}A_{*,-1}^{p+6 \over 2(p+4)}.
\end{equation}
The latter seems to be more realistic and considered here. The radio
afterglow thus will peak when the observation frequency, $\nu_{\rm
obs}$, crosses $\nu_{\rm a}$ at
\begin{equation}
t_{\rm peak} \sim 3\times 10^{6} ~{\rm s}~ ({\nu_{\rm a} \over
8.64~{\rm GHz}})^{-1} \beta^{4p-6 \over p+4} \epsilon_{\rm
e,-1}^{2-p \over p+4}\epsilon_{\rm B,-1}^{p+2 \over
2(p+4)}A_{*,-1}^{p+6 \over 2(p+4)}.
\end{equation}
For typical parameters $\epsilon_{\rm e,-1}\sim 1$, $\epsilon_{\rm
B,-1}\sim 1$, $\beta \sim 0.1$ and $A_* \sim 1$, {\it we expect the
SN radio afterglow will peak at $\sim 100$ days}.

Using Eq. (\ref{eq:F_numax}) the peak flux can be estimated as
\begin{equation}
F_{\rm \nu_{\rm radio}, peak} \sim 1~{\rm Jy}.
\end{equation}
For $\nu_{\rm m}<\nu_{\rm radio}<\nu_{\rm a}<\nu_c$, $F_{\nu_{\rm
radio}} \propto \beta^2 t^{5/2}$. For $\beta \sim $ const., we have
$F_{\nu_{\rm radio}} \propto t^{5/2}$, increasing with time rapidly.
The current two data reported in GCN \cite{Sod08,van08} do suggest a
quick rise of the radio flux and support our assumption $\beta \sim$
const. For $\beta \propto t^{-1/(3+k)}$, we have $F_{\nu_{\rm
radio}} \propto t^{5/2-2/(3+k)}$. As long as the observation
frequency is above $\nu_{\rm a}$, the light curve is described by
Eq.(\ref{eq:LC}).

\section{Conclusion and Discussion}
Transient 080109/SN\,2008D presents the first evidence for a mild-relativistic outburst, $\sim
10^{46}\,{\rm erg}$, preceding the main SN component, thus confirming previous speculation in
SN\,2005bf (Folatelli et al. 2006).

It sets a lower limit of the local event rate of its kind as $1/3{\rm{yr}}/({\rm{0}}{\rm{.0276
Gpc}})^3 \sim 1.6 \times 10^4 {\rm{yr}}^{{\rm{ - 1}}} {\rm{Gpc}}^{{\rm{ - 3}}}$, comparable
with the local rate of Ibc SNe, $\sim 4.8 \times 10^4 {\rm{yr}}^{{\rm{ - 1}}}
{\rm{Gpc}}^{{\rm{ - 3}}}$, and thus indicates a vast majority of X-ray transients have a wide
opening angle of $\gtrsim 100^\circ$. The collimation-corrected energy is of $\sim 5\times
10^{45}$ erg. The wide angle budget, together with the self-consistent interpretation of the
transient and its early afterglow with the on-beam model, largely rules out the off-axis
viewing model for this transient.

The host NGC2770, a spiral galaxy with copious ${\rm H\alpha}$ sign, stands out from the
star-forming dwarf galaxies typically hosting GRBs/XRFs. Transient 080109 puts itself on the
upper border of the nearby GRB/XRF collection in terms of the host metallicity (Berger \&
Soderberg 2008b; Sollerman et al. 2005).

As a result, this event may unveil a continuum from energetic GRB (top-right of Figure 2) to
ordinary Ibc SN (bottom-left of Figure 2).

(1) Whether or not every Ibc SN has a quasi-jet outburst proceeding the main SN component is
still uncertain even the discovery of Transient 080109. For this reason we mark the ordinary
Ibc SN and X-ray transient/SN populations with a dash ellipse in Figure 2.

(2) Soderberg et al. (2006) showed that producing GRBs/XRFs needs a relativistic ejecta
carrying at least $10^{48}$ erg. We show in this {\it Letter} that X-ray transient population
couples $\sim 10^{46}$ erg to relativistic material regarding this found one marks the
transition between GRB/XRF and ordinary Ibc.

(3) While under-luminous GRBs/XRFs are likely powered by moderate-relativistic material, X-ray
transients are likely powered by mild-relativistic material.

(4) GRBs have an average opening angle of $\sim \ 10^\circ$ while a vast majority (if not all)
of X-ray transients have a much wider one of $\sim 100^\circ$. There is a negative correlation
between radiated energy and opening angle from GRB to XRF to X-ray transient.

(5) Materials with higher bulk Lorentz factor tend to have a shallower energy-velocity
distribution leading to spikeful behavior as shown in GRB/XRF prompt lightcurves and
hypernova's broad-lined spectra. Materials with lower bulk Lorentz factor tend to have a
steeper energy-velocity distribution and thus largely couple with each other leading to
spikeless/little-spiked behavior as shown in various optical afterglows. The decay laws in
terms of velocity for each event in Figure 2 (from left to right) matches this principle.

\acknowledgements It's a pleasure to thank D. Watson for providing the X-ray data, D.
Malesani, J. P. U. Fynbo, J. Hjorth, J. Sollerman, G. Leloudas, J. S. Deng for discussion, and
J. Gorosabel, W. D. Li, K. Page, and D. M. Wei for discussion/communication. The Dark
Cosmology Centre is funded by the Danish National Research Foundation (DNRF). YZF is supported
by a postdoctoral grant from DNRF, the National Science Foundation (grant 10673034) of China
and a special grant of Chinese Academy of Sciences. YCZ is supported by National Science
Foundation of China (grant 10703002).

\end{document}